# A Node Embedding Framework for Integration of Similarity-based Drug Combination Prediction

**Liang Yu\*, Mingfei Xia, Lin Gao**

**School of Computer Science and Technology, Xidian University, Xi'an 710071, Shaanxi, China**
**\* To whom correspondence should be addressed: Email: lyu@xidian.edu.cn.**

**Abstract**

**Motivation**: Drug combination is a sensible strategy for disease treatment by improving the efficacy and reducing concomitant side effects. Due to the large number of possible combinations among candidate compounds, exhaustive screening is prohibitive. Currently, a plenty of studies have focused on predicting potential drug combinations. However, these methods are not entirely satisfactory in performance and scalability.

**Results**: In this paper, we proposed a Network Embedding framework in Multiplex Networks (NEMN) to predict synthetic drug combinations. Based on a multiplex drug similarity network, we offered alternative methods to integrate useful information from different aspects and to decide quantitative importance of each network. To explain the feasibility of NEMN, we applied our framework to the data of drug-drug interactions, on which it showed better performance in terms of AUPR and ROC. For Drug combination prediction, we found seven novel drug combinations which have been validated by external sources among the top-ranked predictions of our model.

**Availability**:

**Contact**:

**Supplementary Information**:

## 1 Introduction

Drug combination therapies have been used for the treatment of complex disease such as cancer and Type II diabetes due to the advantage of higher efficacy, fewer side effects and less dose with less toxicity compared to single drug therapies(He, et al., 2016). The main reason is that complex disease is normally involved in multiple pathways and multiple genes, which limits the therapeutic opportunity of one-gene-one-drug. Despite the increasing number of drug combinations in use, many of them were found in clinic or by biological experience. The mechanistic understanding of synergistic drug combinations remains largely elusive, which makes it difficult to propose new drug combinations. Although many approaches have been proposed to seek drug combinations, there are always some inevitable limitations for them.

To promote the development of in-silico methods for computing drug synergy, the Dialogue for Reverse Engineering Assessments and Methods (DREAM) consortium launched an international open challenge for the development of computational models that can be used to objectively and systematically evaluate the accuracy and specificity of drug synergy predictions. Experiments are needed for drugs with gene expression, but many drugs don't have gene expression on all kinds of cell lines. Therefore high-throughput screening seems to be the first choice. However, high-throughput screening method is not suitable to search drug combinations, because the large-scale experiments of possible drug combinations used are very consuming both in time and in money (Cokol, et al., 2011; Winter, et al., 2012) and it's also not feasible to seek synergistic drug combinations.

Some drug combination predictions using protein-protein networks are based on hypothesis that similar drugs can affect similar proteins (Jia, et al., 2009). However, protein-protein network is biased in which some relations superficially exist while pharmacokinetic and pharmacodynamics are ignored. Of course, distance between drug targets in PPI network is an important element in prediction of drug combinations.

Molecular networks, such as signal transduction and gene regulatory networks are also used to predict drug combinations. Åsmund Flobak et al proposed logical model simulations which can be used to automate reasoning on network dynamics even with scarce knowledge of kinetic parameters(Flobak, et al., 2015), and have been used to describe and predict the behavior of molecular networks affected in human disease. However, molecular network is biased based on knowledge and simulated biological metabolism is not accurate.

Alternatively, some computational approaches have been proposed, which use drug similar network weights as a features and train computational models to predict drug combinations(Atias and Sharan, 2011; Cheng and Zhao, 2014; Gottlieb, et al., 2012; Li, et al., 2015; Sridhar, et al., 2016; Vilar, et al., 2013). For example, Li et al proposed PEA to predict drug combinations using seven layer networks. But most of them often have limitations that the feature is fixed and involves only edges in similarity networks .Without consideration of nodes in the multiplex network, we can't fully integrate topological information into the feature.

Drug-drug interactions(DDIs) are classified as pharmacokinetic and pharmacodynamics. Pharmacokinetic interactions are usually associated with an adverse or exaggerated response and pharmacodynamics interactions are implicated in both synergistic and detrimental effects(Sridhar, et al., 2016). Therefore, when the model got better performance in DDIs prediction, it's also suitable for drug combination prediction.

Network embedding has achieved vast success in social network classification and clustering, such as DeepWalk, LINE and Grape. But these methods are only suitable for single-layer network, not applicable to multiplex networks data integration. OhmNet(a hierarchy-aware unsupervised node feature learning approach) need tissue hierarchy interactions. We consider different multiplex networks as sampling from real network in different aspects. But not every network has the same weight.

In this paper, we proposed a framework called NEMN, which uses drug similar networks to predict synergistic drug combinations. Multiplex drug similarity network were built using drug multiple omics data. The novelty of NEMN can be seen in four aspects: (1) Complementary integration of the topology of multiplex networks into data feature. (2) Available to predict relations of drug-disease or drug-target. (3) Easy to scalable for more drug similar networks. (4) Use of different weights of importance when sampling multiplex networks. In consequence, our method got better results compared with INDI (Gottlieb, et al., 2012), PLS (Sridhar, et al., 2016) and mashup (Cho, et al., 2016) when using same drug similarity data from Gottlieb, et al. (2012) to predict drug-drug interactions. In the end, we predicted seven synergistic drug combinations and verified them by literature.

## 2.Materials

For Drug combination prediction, we calculated six similarity networks according to five drug-based as well as one target-based similarity measures, 1284 drugs were preserved after we selected drugs existed in all six drug similarity networks. Drug combinations data were obtained from the Drug Combination Database (DCDB) (Liu, et al., 2010) and the website of food and drug administration (FDA).

### 2.1 Drug-drug similarity network

We defined and computed six drug-drug similarity measures, including the chemical similarity, the similarity based on side effect, the anatomical therapeutic and chemical (ATC) classification system similarity, the similarity based on text-mining, the similarity based on distances in a protein-protein interaction network, and firstly using the categories of drug similarity.

### 2.1.1 Chemical similarity network

We downloaded the data of small molecule drugs with detailed structural formula and molecule from DrugBank (Wishart, et al., 2018). Each vertex of the network was weighted by using molecular fingerprint feature vectors with 1024-dimensions by the calculation of the PaDEL-Descriptor software (Yap, 2011). In the binary vector, elements 1 or 0 indicate the presence or absence of a specific chemical substructure seperately. The molecular fingerprint feature has been widely used in the study of quantitative structure activity relationship (Dimova, et al., 2013; Rabal, et al., 2015). A jaccard function was used to calculate drug similarity based on 1024-dimensions mentioned above.

### 2.1.2 ATC similarity network.

The Anatomical Therapeutic Chemical (ATC) Classification System, which includes 5 different hierarchical levels, was used to classify drugs into different groups according to the organ they acted on and the therapeutic chemical characteristics. The similarity between two ATC codes is derived according to their prior probabilities (frequency) and the probability of their commonality (Zhao and Li, 2010), which is defined as their longest matched prefix:

$$S(i,j) = \frac{2*\log(\Pr(prefix(i,j)))}{\log(\Pr(i)) + \log(\Pr(j))} \quad (1)$$

where $prefix(i,j)$ is the longest matched prefix of ATC code $i$ and $j$. Note that drugs may have more than one ATC code, we then define the maximum ATC code similarity as TS:

$$TS(d_1, d_2) = \underset{i \in ATC(d_1), j \in ATC(d_2)}{MAX}(S(i,j)) \quad (2)$$

where $ATC(d)$ represents all the ATC codes belonging to drug $d$.

**2.1.3 Side-effect similarity network**

Drug side effects were obtained from SIDER 4.1 (http://sideeffects.embl.de/) (Kuhn, et al., 2016), a public resource containing drug side-effect information. We assigned a side-effect profile to each drug in our dataset, whose elements encode for the presence or absence of each of the side-effect by using 1 or 0 respectively. As mentioned above, we defined and computed similarity scores between drugs according to jaccard similarity score between their side-effect profiles.

**2.1.4 Text-mining similarity network**

We downloaded drug-drug similarity score from stitch dataset (http://stitch.embl.de/) based on the times of their cooccurrence in the same paper (Szklarczyk, et al., 2015).

**2.1.5 Target distance similarity network**

The similarity between each pair of drug target proteins in the human PPI network was calculated by using the shortest distance between drug targets. PPI network was downloaded from human protein reference database (HURM) (Keshava Prasad, et al., 2009). We transformed distance to similarity by the formula as fellow.

$$\frac{1}{1+\text{distance}(g_1,g_2)} \quad (3)$$

The similar equals to 1 when distance equals to 0. The similar becomes smaller as the distance becomes larger.

**2.1.6 Category similarity network**

We firstly used the categories of drug to describe information about drugs. The categories of drug reflect the compound and pharmacology of drugs. We assigned a category profile to each drug from DrugBank, whose elements encode for the presence or absence of each of the categories using 1 or 0 respectively. As mentioned above, we defined and computed similarity score between drugs according to jaccard similarity score between their category profiles.

1284 drugs were preserved after we selected drugs exsited in all six drug similarity networks. As a result, six drug similarity networks were homologous and heterogeneous and each layer of network has exactly 1284 nodes.

**2.2 Drug combination data**

We downloaded drug combinations from DCDB dataset and the website of food and drug administration (FDA). 275 pairs of drug combination were obtained from FDA to evaluate the importance of different drug similarity networks. There remained to be 947 pairs of drug combination from DCDB after the removal of those that already existed in FDA. But after overlapping 1284 drugs in similarity networks with those in chosen drug combinations, we finally obtained 239 and 275 pairs of drug combination respectively from DCDB and FDA as positive samples.

**3 Methods**

**3.1 evaluate network importance**

We think drug similarity networks reflected drug similarity from different aspects, each of network had different information volume and absolutely should have different weights to represent network importance. So, we used a part of data, such as, drug combination or DDIs to evaluate the importance of each network. For each of network, we found that the edge weight between known drug combinations in networks was higher than that between randomly selected ones when using Wilcoxon rank-sum statistic to sample 1,000,000 instances. So six drug similarity networks had information about drugs and each of them had different topological structure. The weight of synergistic drug combinations is larger than that of randomly sampled edges, which has been proved by Zhao, et al. (2011). It's approved that the network in which the edge weights of known drug combinations have greater significance than in other networks has more and accurater information. We used p-values calculated by Wilcoxon rank-sum statistic to weight each network in the form of a decreasing function. The weight of the network is used as the times of sampling from the according network.

According to the algorithm of DeepWalk (Perozzi, et al., 2014) and Node2Vec(Grover and Leskovec, 2016), the weights of the network usually set 10 and we also followed this chosen. We considered 10 as the mean of weights of networks and slightly modified the weight according the p-value calculated by Wilcoxon rank-sum statistic. It's reasonable when the smaller p-value compared with the bigger weight, because the weight and p-value is relative to other weight and p-value. According to the paper of Grover, et al(Grover and Leskovec, 2016)., weights should be about 10. After experiment, we think the weight is not sensitive to the result but should not be too big or too small.

**3.2 network information sampling**

We achieved this by utilizing a flexible biased random walk procedure that can explore neighborhoods to sample drug similarity networks.

### 3.2.1 Random Walks

Formally, given a source node $u$, we simulate a random walk of fixed length $L$. Let $C_i$ denote the $i$ th node in the walk, starting with $C_0 = u$. Nodes $C_i$ are generated by the following distribution:

$$P(C_i = x \mid C_{i-1} = v) = \begin{cases} W_{vx} & if\ (v,x) \in E \\ 0 & otherwise \end{cases} \quad (4)$$

Where $W_{vx}$ is the normalized transition probability between node $v$ and node $x$ (Grover and Leskovec, 2016).

### 3.2.2 Biased random walk

Random walk does not allow us to account for the network structure and guide our search procedure to explore different types of network neighborhoods. We used a 2nd order random walk with two parameters $p$ and $q$ refer to node2vec (Grover and Leskovec, 2016) with a piecewise functions which guide the walk as following : Consider a random walk that just traverses edge (t, v) and resides at node v. The walk now needs to decide on the next step so it evaluates the transition probabilities $W_{v,x}$ on edges (v, x) leading from v.

$$\alpha(t,x) = \begin{cases} 1/p & if\ d_{tx} = 0 \\ 1 & if\ d_{tx} = 1 \\ 1/q & if\ d_{tx} = 2 \end{cases} \quad (5)$$

We set the unnormalized transition probability to $W_{vx} = \alpha(t,x)W_{vx}$, where $d_{tx}$ denotes the shortest path distance between nodes $t$ and $x$. Note that $d_{tx}$ must be one of {0, 1, 2}, and hence the two parameters are necessary and sufficient to guide the walk. Finally, we used biased random walk to acquire many paths.

### 3.3 FEATURE LEARNING

Random walk sequences are considered as sentences in natural language processing in which a node equals to a word, so we used Word2Vec method provide by Google(Le and Mikolov, 2014).

### 3.3.1 SkipGram

SkipGram (Le and Mikolov, 2014; Mikolov, et al., 2013) is a language model that maximizes the cooccurrence probability among the words that appear within a window ($k$) in sequences.

SkipGram tries to maximize classification of a node based on another node in the same path. More precisely, we use each current node as an input to a log-linear classifier with continuous projection layer and then predict nodes within a certain range before and after the current node in the biased random walk path. We found that increasing the range improves quality of the resulting node vectors, but it also increases the computational complexity. Since the more distant words are usually less related to the current node than those close to it, we give less weight to the distant node by sampling less from those nodes in our training examples. We need to sure the distance of window in the biased random walk path. Where we set $k = 5$(see appendix), all possible collocations in random walk that appear within the window $k$ were considered as positive sample and negative sample using negative sample (Mikolov, et al., 2013) to generated. Given the representation of $V_j$, we would like to maximize the probability of its neighbors in the walk.

---

**Algorithm** $SkipGram(f, P, k, d)$

**Input: windows size $k$**
  **random walk path $P$**
  **embedding size $d$**
**Output: matrix of vertex representations** $\phi \in \mathbf{R}^{|V|*d}$
**1: for each $v_j \in P$ do**
**2:   for each $u_k \in P[j-w:j+w]$ do**
**3:     $J(\phi) = \log \Pr(u_k \mid \phi(v_j))$**
**4:     $\phi = \phi - \alpha * \dfrac{\partial J}{\partial \phi}$**
**5:   end for**
**6: end for**

---

$P[j-w:j+w]$ represented nodes appear in random walk path has distance smaller than the window size of $k$. The function of $\mathbf{Pr}$ is equal to logistic classifier. It's also suitable for formula 6 and 7.

### 3.3.2 Optimization

The optimization phase is made efficient using asynchronous stochastic gradient (SGD) (Bottou, 1991).

The training objective of the Skip-gram model is to find node representations that are useful for predicting the surrounding nodes in a random walk path. More formally, given a lots of random walk paths training nodes $v_1$, $v_2, v_3, \ldots, v_L$, the objective of the Skip-gram model is to maximize the average log probability

$$\max(\frac{1}{L}\sum_{t=1}^{L}\sum_{-c \leq j \leq c, j \neq 0} \log Pr(v_{t+j} \mid v_t)) \quad (6)$$

$$\Pr(\{v_{i-w},...,v_{i+w}\} \setminus v_i \mid \phi(v_i)) = \prod_{\substack{j=i-w \\ j \neq i}}^{i+w} \Pr(v_j \mid \phi(v_i)) \quad (7)$$

The basic Skip-gram formulation defines using the softmax function as fellow.

$$p(v_O \mid v_I) = \frac{\exp(d'_{v_O}{}^T d_{v_I})}{\sum_{w=1}^{N} \exp(d'_{v_w}{}^T d_{v_I})} \quad (8)$$

where $d_v$ and $d'_v$ are the input and output vector representations of $v$ and $N$ is the number of nodes in the Nodes.

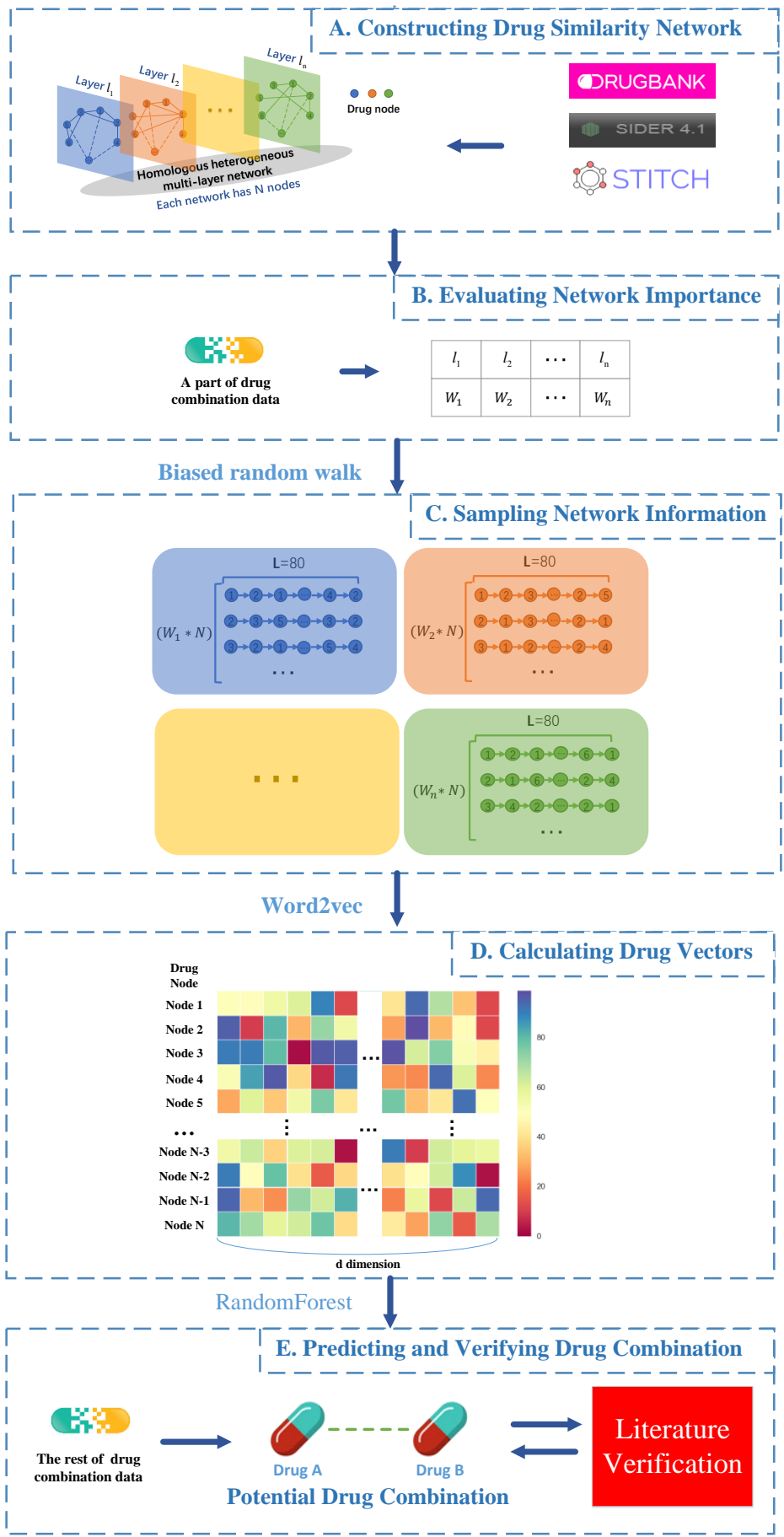

Fig 1.Workflow for NEMN. Multiplex homologous heterogeneous drug similarity networks were built on DrugBank, SIDER and STITCH database, and each network has N drug node. The result is N dimension importance vectors by using a part of drug combination data to evaluate the importance of each of drug similarity network. A biased random walk method was adopt to sampling network to obtain random walk paths, and every network got different number of paths. Drug node vectors and classifier were combined to predict drug combination.

# 4 Results

NEMN got better performance compared with the baseline function of INDI and Probabilistic Soft Logic(PSL) model, mashup, and two variant of NEMN in DDIs predict task. After that, NEMN was tenstedin Drug Combination Predication and analysis the result of drug combination predicted.
The paragraph should be reorganized. You can say:
We first test the performance of NEMN in *** task, by comparing with ***. The results showed ***. Then we test NEMN in *** task, by comparing with ***. The results showed ***.

## 4.1 evaluate NEMN by using other data and in comparison with other methods

We used drug-drug interactions and drug similarity from Gottlieb, et al. (2012), the dataset contains two types of interactions: (1) CYP-related interactions (CRDs), (2) non-CYP related interactions (NCRDs), including 10,106 CRD and 45,737 NCRD DDIs across 807 drugs. Meanwhile, Seven drug similarity networks(chemical-based, Ligand-based, side-effect-based, annotation-based target-sequence-based, target-PPI network-based, target-Gene Ontology-based ) are also provided by Gottlieb, et al. (2012)

INDI was often used as the baseline for DDIs prediction, based on seven drug similarity network, and 49 features were constructed by INDI(). The Probabilistic Soft Logic(PSL) model was used for DDIs prediction and got better performance when compared with INDI using same data. Mashup is a feature learning algorithm that learns low-dimensional representations for nodes based on their steady-state topological positions in the networks. It's similar to NEMN, also got nodes feature for prediction.

Two variant of NEMN also was taken into account for committed to validate the novelty of the method of NEMN. We designed NB_NEMN which used random walk rather than biased random walk and NS_NEMN which gave same weights for each network rather than evaluate the importance of each network.

. Firstly, we randomly selected 20% interactions from CRD interactions or NCRD interactions with edge weights equal to 1. After that, we evaluated network importance by Wilcoxon rank-sum statistic to compare the elaborately selected 20% interactions with randomly selected edges from the same similarity network. Tabel-1 showed weights and p-values by Wilcoxon rank-sum test. It has been confirmed by DeepWalk and node2vec that the weights for each network should be about 10.

We used biased random walks of length 80 to sample networks by using $p=1$ and $q=3$(equals to node2vec). The weights calculated above used to ensure the times of sampling from node in each similarity network. We got many paths including every drug similarity network information and network topology. Drug node vector was easy to get by using Word2vec model which can provide many biased random walk paths. We discussed the result of prediction using biased random walk and random walk later.

Table 1.Weights of drug similarity network and p-value using Wilcoxon rank-sum test in drug-drug interactions data.

|  | CRD | | NCRD | |
| --- | --- | --- | --- | --- |
|  | p-value | weights | p-value | weights |
| ATC | 1.7e-5 | 8 | 2.5e-275 | 12 |
| Chemical | 6.8e-05 | 10 | 0 | 15 |
| Distance | 2.2e-06 | 11 | 0 | 15 |
| GO | 5.9e-2 | 5 | 0 | 15 |
| Ligand | 2.8e-4 | 7 | 1.8e-66 | 8 |
| Seqs | 7.6e-46 | 15 | 0 | 15 |
| Sise Effect | 4.6-09 | 13 | 1.2e-57 | 8 |

A random forest classifier was adopted to classify the CRD interactions and NCRD interactions. The selection of parameters in the model followed supplementary materials. NEMN, mashup, NS_NEMN and NB_NEMN used the same parameters in random forest classifier to predict interactions. We not directly considered splice two drug node vector as feature to classify, because two different features represent the same of drug combination wasn't reasonable. So, we use Euclidean distance in every dimension of the feature. Finally, 50 dimension vector was used to represent drug combination. 10-fold cross-validation was used for the evaluation. We applied the six methods to each fold and reported average and standard deviations of our chosen metrics for each one. Tables 1–2 presented average and standard deviations for recall, AUPR and AUC in cross-validation experiments.

From Table 2(right ?), we can see NEMN significantly outperformed both baselines in AUC, AUPR and F1-score when experimenting on two types of interactions. For AUPR in the best case, NEMN improved up to 62% over the PSL in which it ranged from 0.34 to 0.91. Mashup also got high AUC, AUPR and recall, which indicated the method exploiting node embedding may be more effective in predication when using multiplex network data. And compared with mashup, NEMN also got better performance.

Table 2. Average AUPR, AUC and F1 scores, and standard deviation for 10-fold CV for CRD interactions

| Method | AUPR | AUROC | F1 |
| --- | --- | --- | --- |
| INDI | 0.15+/-0. 007 | 0.92+/-0.003 | 0.24 +/- 0.005 |
| PSL | 0.34+/- 0.02 | 0.96+/-0.003 | 0.40 +/- 0.02 |
| NEMN | 0.918+/-0.01 | 0.979+/-0.01 | 0.944 +/- 0.01 |
| Mashup | 0.92+/-0.01 | 0.977+/-0.01 | 0.94 +/- 0.01 |
| NS_NEMN | 0.918+/-0.01 | 0.977+/-0.01 | 0.942 +/- 0.01 |
| NB_NEMN | 0.916+/-0.01 | 0.976 +/- 0.01 | 0.941 +/- 0.01 |

Table 3. Average AUPR, AUC and F1 scores, and standard deviation for 10-

fold CV for NCRD interactions

| Method | AUPR | AUROC | F1 |
| --- | --- | --- | --- |
| INDI | 0.64 +/- 0.01 | 0.95 +/- 0.003 | 0.63 +/- 0.01 |
| PSL | 0.78 +/- 0.02 | 0.97 +/- 0.003 | 0.70 +/- 0.02 |
| NEMN | 0.964+/-0.01 | 0.993+/-0.01 | 0.943+/-0.01 |
| mashup | 0.962+/-0.01 | 0.993+/-0.01 | 0.938 +/-0.01 |
| NS_NEMN | 0.962+/-0.01 | 0.992+/-0.01 | 0.943+/-0.01 |
| NB_NEMN | 0.963+/-0.01 | 0.992+/-0.01 | 0.941+/-0.01 |

On the whole, Feature learning methods more suitable for DDIs prediction when used multiplex network as input. Mashup and NEME significantly outperformed INDI and PSL in AUC, AUPR and F1-score in CRD and NCRD interactions. For AUPR in the best case in CRD interactions, the two models improved up to 62% and 83% over PSL and INDI respectively. For AUC, the two models had comparative accuracy with PSL. For F1-score, NEME improved nearly 20% over PSL. Interestingly, the three different kinds of NEME and mashup method always performed better than INDI and PSL. This finding demonstrated the effectiveness in combining multiple similarities of NEMN. As for Table2, the result is similar to Table1, NEMN outperformed PSL and INDI considering AUPR and F1-score.

To assess the benefit of using biased random walk and the meaning of evaluating network importance, we also compared NEMN model with NS_NEMN and NB_NEMN. It is shown that biased random walk can lead to better results.

**4.2 Drug Combination**

Drug combinations were obtained from section 2.2, and drug similarities from section 2.1. After that, we got 514 pairs of drug combination among which 275 were from FDA and 239 from DCDB. Drug similarity networks were homologous networks, each of which were built based on different similarity measures and contained the same 1287 nodes.

Firstly, we evaluated network importance by using drug combinations from FDA. By mapping drug combinations from FDA to each drug similarity network, we got an edge list for each network. We compared edge weights of drug combinations with that of randomly selected edges from networks and got p-values by using Wilcoxon rank-sum statistic. We found that drug combinations has greater weights compared with randomly selected edges, which agreed with the conclusion by Zhao, et al(Zhao, et al., 2011). So we assigned the similarity network with higher weights when its according p-value was smaller. Tabel-4 showed weights and p-values by Wilcoxon rank-sum test in each drug similarity network.

Table 4. Weights of drug similarity network and p-value using Wilcoxon rank-sum test in combination data.

| Network | p-value | weights |
| --- | --- | --- |
| ATC | 2.90153447482e-29 | 15 |
| Indication | 3.69223052984e-22 | 13 |
| Text-mining | 9.87695010037e-36 | 18 |
| Side-effect | 4.00275686413e-10 | 10 |
| chemical | 0.0792944676514 | 5 |
| Drug-target distance | 2.29044980103e-05 | 8 |

Secondly, we used biased random walks of length 80 to sample networks by using p=1 and q=3(equals to node2vec). The weights calculated above used to sure the times of sampling from node in each similarity network. We got many paths including every drug similarity network information and network topology.

Using Word2vec algorithm to calculated drug node vector. Mang parameters in word2vec, we selected same parameters with node2vec, and output size we set to 5, due to we had 416 pairs drug combination as positive. high dimension lead to overfitting. Other drug combination predict methods has many features and just has hundreds of drug combination pairs. Overfitting is a potentially problem in drug combination prediction due to little drug combination data especially in using multi-layer machine learning to drug combination prediction.

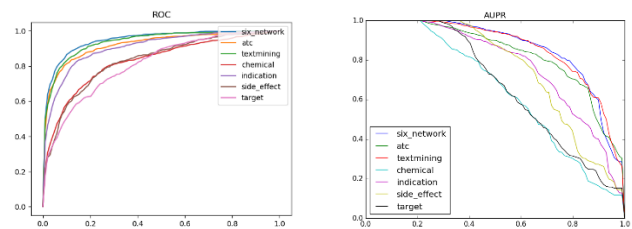

Fig. 2: Receiver Operating Characteristic (ROC) and Precision-recall curves in Drug combination Prediction.

We calculated ROC and AUPR by using different network data. To quantitatively assess the performances of the NEMN model with all six features or each single feature in predicting effective drug combinations, we used the 416 drug combination pairs as positive of our gold standard and negative samples that random selected four times of the size of positive from all drug pairs. A 10-fold cross validation accompanied with the receiver operating characteristic (ROC) curve analysis. As a result, the model with whole features (AUC=0.95) exhibits better performance than those with single feature (AUC=0.81–0.94) (Fig. 2). Among the six features, textmining has the most predicting performance (AUC=0.94). For AUPR, six features was also got more performance of 0.86. It turns out that the whole-feature NEMN model performance better than single feature.

To validate the reliability of our method, we further check whether the predicted drug pairs were validated in external literatures which were not used to build the training dataset for the NENG model. Due to negative samples are unknown, we every train model using different negative samples that random selected from all drug pairs. So, we trained three classifiers using same positive sample and different negative sample. The results we predicted drug combinations meanwhile exist in three different classifiers predicted results and top-300 in every classifier. We chose random forest classifier to train model. The result of drug combination index sort by the sum of different classifier index. Finally, we think sort in top and not in positive sample are predicted by NEMN model.

### 4.3 External literature validation

To validate the reliability of our method, we further check whether the predicted drug pairs were validated in external literatures which were not used to build the training dataset for the NENG model. (See Table5).

We rank these new predictions and consider the top 30 interactions and not contain in train data as shown in Table-5, it's our prediction. So, we need to validate in literature.

Table 5. Top ranked NEMN model predictions for interaction unknown in drug combination.

| ALL Rank | DrugBank_ID | DrugBank_ID | Predict Rank | Pubmed ID |
|---|---|---|---|---|
| 4 | DB00515 | DB01229 | 1 | 12376205 |
| 7 | DB00232 | DB00999 | 2 | |
| 12 | DB00790 | DB00999 | 3 | 9048272 |
| 22 | DB00530 | DB01101 | 4 | 23393373 |
| 27 | DB00530 | DB01229 | 5 | 20332457 |
| 30 | DB00999 | DB04861 | 6 | 20556921 |

The top predicted drug combination is between DB00515 and DB001229 corresponding to Cisplatin and Paclitaxel, both cisplatin/paclitaxel regimens showed excellent activity with manageable toxicity in patients with advanced ovarian cancer has proved by De Jongh F E et al with a randomised phase I/II trial in 49 patients with ovarian cancer. Another evidence is Gynecological Oncology Group trial comparing cisplatin and cyclophosphamide versus cisplatin and paclitaxel. The response rate for the cisplatin and paclitaxel combination was 77%, with a median survival that was 13.1 months longer than that of the cisplatin and cyclophosphamide-treated group. Based on this trial, cisplatin and paclitaxel became the standard first-line treatment regimen for patients with advanced ovarian cancer(de Jongh, et al., 2002).

For the combination of DB00232 and DB00999 hasn't been proved by any literature. But the target of DB00232 were Q13621, P00915, P00918 and P22748 has two same target of DB00999. In the paper of Zhao, et al. (2011) found among 281 such pairs of drugs has proved by FDA, 100 share target proteins - a significantly higher proportion than expected by chance.(p-value of $10^{-5}$, Fisher's exact test) So, DB00232 and DB00999 maybe has synerstic effect.

All of these drug combination pairs has proved by experiments in patients, and it's the most reasonable way to explanated drug combination result was accuracy.

## 5 Discussion

Developing combinatorial therapies for complex disease treatment has attracted increasing attention due to their great potentials as compared to monotherapies. One major challenge to develop combinatorial therapies is the large search space of possible combinations, therefore computationally predicting drug combination effects and prioritizing drug combinations is vitally important. In this paper, we have proposed a computational framework NEMN to predict drug combination pairs.

In drug-drug interactions prediction, we observed that NEMN has better performance than pervious work like INDI and PSL, and method based on node embedding such as mashup also has high AUC and AUPR. By construct multi-layer drug similarity network to description drug interaction. Previous drug combination work has fixed feature that similarity network edge weights, it's can't combine network topology information with drug similarity to feature. So, feature is limited, can't using feature to description drug interaction accurately.

A framework based on network embedding were proposed by ours to dell with multi-layer homologous heterogeneous node vector representation. Two novelty of NEMN are using biased random walk and evaluate network importance has proved in Drug-Drug interactions prediction task. We sample multi-layer networks using biased random walk and evaluate each of network has different weights got better performance. Due to NEMN is data driven, not based on any hypothesis, we can use NEMN method in all kinds of situation, only input are multi-layer network or single-layer network and need a bit of prediction interactions.

For drug combination prediction, the top-30 ranker in NEMN method has been validate in literature, and all of them got well performance in biology experiment especially in Phase I or II in Clinical cancer patients. There are several directions worth explorations in future studies. First, We proposed framework not only drug-drug interactions prediction, but also for all interactions prediction. Such as Drug-Target interactions prediction, Drug-Disease interactions prediction. Second, we compared single-layer network with multi-layer network results ,in which indicate multi-layer network better proformance than single-layer network. We just using six drug similarity networks, many others drug similarity network can be take including GO information and pathway information into account. Different aspects information for drug can accuracy description drug, NEMN method also can using a more reasonable vector to represent drug node.

Of course, some parameters should be talk feature, Zeng at al proposed to prediction drug-target interaction when using 500 dimension vector to represent drug node and 100-dimension vector to represent target node when just using 100*2000 drug-target interaction(Luo, et al., 2017). Higher dimension usually reflect overfitting. A Fatal weakness of network embedding is that not evaluate the vector of node reasonability. Node vector as feature to classify interactions can't give a reasonable biological explanation. It's also the weakness of NEMN.